\documentstyle[12pt]{article}
 
\begin{document}

\baselineskip=14pt plus 0.2pt minus 0.2pt
\lineskip=14pt plus 0.2pt minus 0.2pt

\begin{flushright}
~~ \\ %atom-ph/9605010 \\
LA-UR-96-426 \\
\end{flushright}

\begin{center}
\large{\bf Analytic Description 
of the Motion of a Trapped Ion in an Even or Odd Coherent State} 

\vspace{0.25in}

\bigskip

 Michael Martin Nieto\footnote
{Email: mmn@pion.lanl.gov }

\vspace{0.3in}

{\it
Theoretical Division\\
Los Alamos National Laboratory\\
University of California\\
Los Alamos, New Mexico 87545, U.S.A.\\
\vspace{.25in}
and\\
\vspace{.25in}
Abteilung f\"ur Quantenphysik\\
Universit\"at Ulm\\
D-89069 Ulm, GERMANY\\}

\vspace{0.3in}

{ABSTRACT}

\end{center}

\begin{quotation}

%*******************************************
\baselineskip=0.333in
%*******************************************

A completely analytic description 
is given of the motion of a trapped ion which is in either an even or 
an odd coherent state.  Comparison to recent theoretical and experimental work is made. 


\vspace{0.25in}

\end{quotation}

\vspace{0.3in}

\newpage

%******************************************************************************
\baselineskip=.33in
%******************************************************************************

Recently,    
Matos Filho and Vogel \cite{vogel} gave an analysis of a trapped ion
\cite{wineland}
which, to very high precision, is
in an even (+) or odd (-) coherent state 
\cite{manko}:
\begin{equation}
\psi_{+}\ = [\cosh{|\alpha|^2}]^{-1/2}
\sum_{n=0}^{\infty}\frac{\alpha^{2n}}{\sqrt{(2n)!}}|2n\rangle ~,
\end{equation}
\begin{equation}
\psi_{-}\ = [\sinh{|\alpha|^2}]^{-1/2}
\sum_{n=0}^{\infty}\frac{\alpha^{2n+1}}{\sqrt{(2n+1)!}}|2n+1\rangle ~.
\end{equation}
Specifically, they gave lovely three-dimensional numerical graphs  of the 
probability densities and Wigner functions, for the 
even and odd cases, as functions of position and time, for 
particular values of $\alpha = \alpha_1 + i \alpha_2$.

Here it is noted  that  closed-form expressions can be 
given for these wave functions, and hence for the probability densities 
and the Wigner functions.  A direct way 
to evaluate the above sums 
\cite{ntprl}
is to 
use generating function techniques 
\cite{ntpla},
yielding 
\begin{equation}
\psi_{+}=
\left[\frac{e^{-\alpha^2}}{\pi^{1/2}\cosh{|\alpha|^2}}\right]^{1/2}
e^{-x^2/2}\cosh(\sqrt{2}\alpha x)~,
\end{equation}
\begin{equation}
\psi_{-}=
\left[\frac{e^{-\alpha^2}}{\pi^{1/2}\sinh{|\alpha|^2}}\right]^{1/2}
e^{-x^2/2}\sinh(\sqrt{2}\alpha x)~.
\end{equation}
(One could, of course, obtain these expressions by other methods 
\cite{manko, vogel2, wolfgang}.)  
Physical intuition is satisfied when 
the above expressions  are transformed to 
two Gaussians displaced on opposite sides of the origin 
\cite{wolfgang}.  Ignoring $e^{-i4\alpha_1 \alpha_2}$, 
\begin{equation}
\psi_{\pm}= 
\left[2 \pi^{1/2}(1 \pm e^{-2 |\alpha|^2})\right]^{-1/2} 
     \left[e^{-(x-\sqrt{2}\alpha_1)^2/2 + i\sqrt{2} \alpha_2 x} 
     \pm e^{-(x+\sqrt{2}\alpha_1)^2/2 - i\sqrt{2} \alpha_2 x}\right]~.
\end{equation}

Time-dependence can be included by letting $\alpha \rightarrow 
\alpha \exp[-i\omega t]$.  We  now take the convention 
$\alpha \rightarrow \alpha_0$ is real, as was done in 
Ref. 
\cite{vogel}.  
Then it is straighforward to calculate additional quantities.  
For example, the probability densities are 
\begin{equation}
\rho_{+} =   \frac{e^{\alpha_0^2[\sin^2\omega t - \cos^2 \omega t]}}
              {\pi^{1/2} [e^{\alpha_0^2} +  e^{-\alpha_0^2}]}
           e^{-x^2}
         [\cosh\{2\sqrt{2}\alpha_0 (\cos\omega t)x\}
            + \cos\{2\sqrt{2}\alpha_0 (\sin\omega t)x\}],
\end{equation}
\begin{equation}
\rho_{-} =   \frac{e^{\alpha_0^2[\sin^2\omega t - \cos^2 \omega t]}}
              {\pi^{1/2} [e^{\alpha_0^2} -  e^{-\alpha_0^2}]}
           e^{-x^2}
         [\cosh\{2\sqrt{2}\alpha_0 (\cos\omega t)x\}
            - \cos\{2\sqrt{2}\alpha_0 (\sin\omega t)x\}].
\end{equation}

The above functions $\rho_{+}$ and $\rho_{+}$
describe the forms of Figs. 1 and 4 in Ref. 
\cite{vogel}.  We show them in our Figs 1 and 2.  Note in particular
that the terms $\exp[-x^2] \times \cosh$  describe the two 
``wave-packets" on opposite sides of the origin.  The $\cos$ 
terms describe the 
interference effects near $x=0$ at $t=(2j+1)\pi/2$.   

These even and odd coherent states have recently been created by 
Wineland's group \cite{wineland}.  They took a trapped $^9Be^+$ ion
and first cooled
it to its zero-point energy.  Then they created the  even/odd 
states by a series of laser pulses that entangled the electronic and motional states of the ion.  The separate packets were separated by as much 
as 800 \AA with their individual sizes at 70 \AA.  

This same group has produced squeezed ground states \cite{wine2}.
Therefore, it is possible that squeezed even and odd states \cite{n}
may be produced in the future.

I gratefully acknowledge conversations  held at the Humboldt 
Foundation Workshop on Current Problems in Quantum Optics, organized by 
Prof. Harry Paul.  This work was supported by the U.S. Department of 
Energy and the Alexander von Humboldt Foundation.  

\vspace{0.25in}

\newpage
\large
\noindent{{\bf Figure Captions}}
\normalsize
%******************************************************************************
\baselineskip=.33in
%******************************************************************************

Figure 1.  A three-dimensional plot of the even-coherent-state probability density, $\rho_+$, as a function of position, $x$,
and time, $t$, for $\alpha_0 = 2$.

Figure 2.  A three-dimensional plot of the odd-coherent-state 
probability density, $\rho_-$, as a function of position, $x$,
and time, $t$, for $\alpha_0 = 5^{1/2}$. 

%******************************************************************************
\baselineskip=.33in
%******************************************************************************

\end{document}